# Multi-Classification of Brain Tumor Images Using Transfer Learning Based Deep Neural Network

Pramit Dutta[1], Khaleda Akhter Sathi[2] and Md. Saiful Islam[3]

[1,2,3] Department of Electronics and Telecommunication Engineering
Chittagong University of Engineering and Technology
Chittagong-4349, Bangladesh
{pramitduttaanik, sathi.ruet19, saiful05eee}@gmail.com

**Abstract.** In recent advancement towards computer based diagnostics system, the classification of brain tumor images is a challenging task. This paper mainly focuses on elevating the classification accuracy of brain tumor images with transfer learning based deep neural network. The classification approach is started with the image augmentation operation including rotation, zoom, horizontal flip, width shift, height shift, and shear to increase the diversity in image datasets. Then the general features of the input brain tumor images are extracted based on a pre-trained transfer learning method comprised of Inception-v3. Finally, the deep neural network with 4 customized layers is employed for classifying the brain tumors in most frequent brain tumor types as meningioma, glioma, and pituitary. The proposed model acquires an effective performance with an overall accuracy of 96.25% which is much improved than some existing multi-classification methods. Whereas, the fine-tuning of hyper-parameters and inclusion of customized DNN with the Inception-v3 model results in an improvement of the classification accuracy.

## 1 Introduction

A brain tumor is defined as an uncontrolled and unnatural growth of neural cells. According to the world cancer report, in this year around 18,020 adults may die from primary brain cancerous disease [1]. Therefore, the early classification of brain tumors into their particular types plays an imperative role to treat the tumor efficiently for reducing the human death rate. In this case, the implication of deep learning methods to classify the tumor images can accelerate the treatment process more effectively. Recently numerous researches have been conducted on deep learning based classification method to increase the classification accuracy of the brain tumor images. For instance, Sajjad et al. [2] employed an unsupervised learning method called the convolutional neural network (CNN) algorithm for the classification of the brain tumor images in different classes. The accuracy was found to be almost 94.58% for the classification of the multiple categories of tumors. Moreover, Amin et al. [3] employed a fusion process using the discrete wavelet transform (DWT) method to extract a more informative tumor region. Then the noise removal process was applied based on a partial differential diffusion filter (PDDF) before segmentation. After that, the CNN model was utilized for classifying the tumors as cancerous and non-cancerous. In another study, Sultan et al. [4] developed an approach for the multi-classification of brain tumors. The classification of the brain tumors images in multiple classes was based on the CNN algorithm. The accuracy was found to be almost 96.13% for the classification of the multiple categories of tumors. Abiwinanda et al. [5] proposed a classification model based on CNN for the multi-classification of brain MRI images. The architecture of the CNN model consisted of different layers i.e. convolution, max-pooling, flattening, and fully connected one hidden layer. The classification method based on CNN provided a classification accuracy of approximately 94.68%.

The major contribution of this paper is to develop an efficient classification method using CNN with the aid of the inception-v3 transfer learning method. Moreover, a suitable learning rate and batch size are employed to make the designed model robust and also faster and smooth the training process. In addition, the effectiveness of the designed method is analyzed by comparing it with some existing classification methods in terms of classification accuracy.

The rest of the paper is organized as follows: Section 2 provides the methodology of the proposed system with a detailed explanation of each of the steps. Section 3 represents the results with analysis and the comparative study is also conducted in this Section. Finally, section 4 shows concluding remarks.

## 2      Proposed Multi-Classification Approach

The process of the proposed multi-classification method is divided into three steps as illustrated in Fig 1. In the first step, the image preprocessing is performed using the image augmentation technique to increase the amount of total image dataset. After the preprocessing step, the image features are automatically generated by employing the Inception-v3 transfer learning method. Finally, the extracted features are feed into the modified DNN to classify the three brain tumor types.

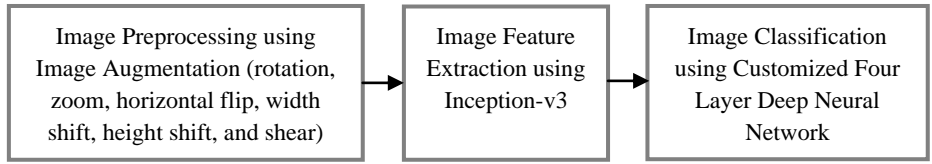

**Fig. 1.** The process of proposed classification method.

### 2.1      Tumor Dataset

The dataset used for this model is obtained from 233 patients with three categories of brain tumor images at different slices (a) 994 axial images, (b) 1045 coronal images, and (c) 1025 sagittal images. This T1-weighted contrast-enhanced image dataset is provided by Cheng [6]. It comprised of 3064 brain tumor MRI images. The datasets are formulated with 1426 meningioma images, 708 glioma images, and 930 pituitary images. Each of the images has a size of 512 × 512 in pixels. For decreasing the computational time and dimensionality the images are resized to 150 × 150 pixels. Then the total image datasets are splitting into training (80% of the total dataset) and validation (20% of the total dataset) dataset. The validation dataset is employed to estimate the proposed classifier model. Moreover, the details explanation of each step of the classifier model are described as following-

## 2.2 Image Augmentation

Before feeding the classifier model, the image augmentation process is performed because of the lower number of the image dataset. For this case, various operation including rotation, zoom, horizontal flip, width shift, height shift, and shear is performed to enhance the diversity of the brain tumor dataset.

## 2.3 Classifier Model using Inception-v3 and DNN

The architecture of the proposed classifier model is comprised of Inception-v3and customized deep neural network as shown in Fig. 2(a) to classify the brain tumor classes effectively.

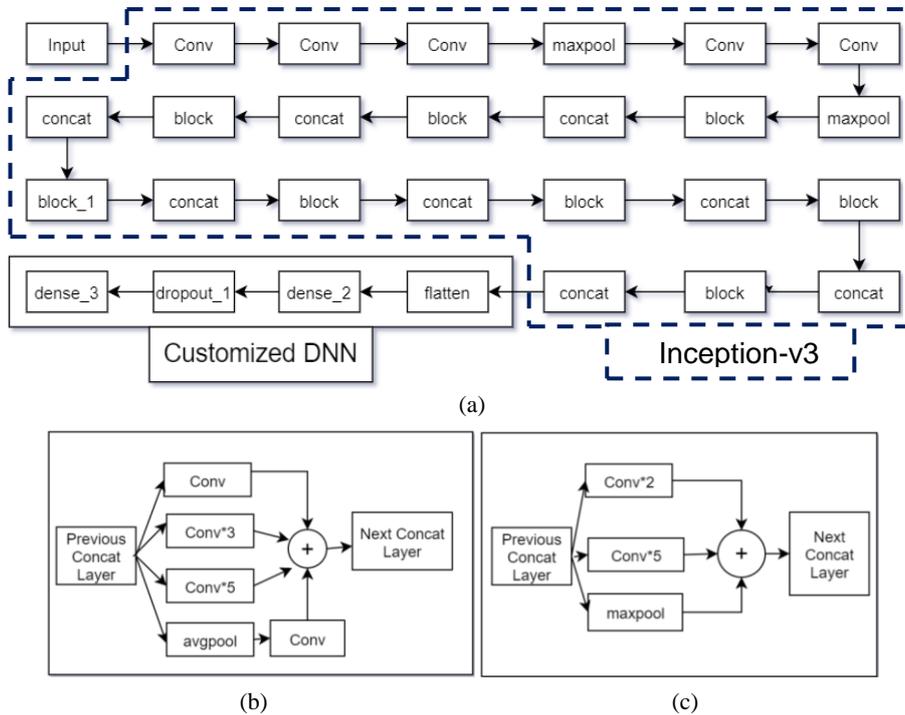

**Fig. 2.** (a) The architecture of proposed classification model, (b) The detail of block, and (c) The detail of block_1.

A. *Feature Extraction using Inception-v3*

After image augmentation, the features are automatically extracted by using the Inception-v3 based transfer learning model. The model utilizes two or three layers of a small convolutional layer based on factorized convolution operation instead of a large convolution layer that reduces the parameter without reducing the efficiency of the model. The factorization process is represented by the block as shown in fig 2(a). This model also employed a grid size reduction technique for mapping the features as shown in fig 2(b). For this factorization process, the 42 layers Inception-v3 model with fewer parameters is much more efficient than the VGGNet [9]. Moreover, the default weights are used for this particular model. This model contains many filters to detect simple feature that is very effective for classification problem. For this reason, the images are convoluted to extract the desired features.

B. *Classification using DNN*

In order to make the classifier more effective, the output of the Inception-v3 model is coordinated with the customized DNN that results in fine tuning of the transfer learning model. Fig.3 shows the network architecture of DNN with four layers where the first layer is used to attend the output of the inception model and the second layer consisting of 1024 neurons and the next layer is utilized to dropout. Finally, the output layer is designed to generate the output with 3 neurons that represents the brain tumor types of glioma, meningioma, and pituitary respectively.

```
flatten (Flatten)            (None, 11520)         0            mixed8[0][0]
____________________________________________________________________________
dense (Dense)                (None, 1024)          11797504     flatten[0][0]
____________________________________________________________________________
dropout (Dropout)            (None, 1024)          0            dense[0][0]
____________________________________________________________________________
dense_1 (Dense)              (None, 3)             3075         dropout[0][0]
============================================================================
Total params: 22,475,427
Trainable params: 22,454,051
Non-trainable params: 21,376
```

**Fig. 3.** The architecture of Deep Neural Network.

For training the network, the Adam is employed as an optimizer and categorical cross-entropy is utilized as a loss function with the learning rate of 0.00003 and batch size of 32. The training process of the designed model is performed in the Google Colab environment using the Graphics Processing Unit (GPU). The training process takes about 6 minutes for 19 epochs (callback stop) with 20 seconds per epoch. During the training, the model occupied 3.06 GB RAM and 2.53 GB GPU in the colab environment.

## 3 Results Analysis

After completing the training process, the loss and accuracy of the designed classifier model are evaluated at different epochs as shown in Fig. 4. The training and validation loss are found 0.0614 and 0.1468 that ensure optimum performance of the classifier model since the validation loss shows a decreasing outline with the increasing of the number of epochs. Moreover, the accuracy is obtained 97.80% and 96.25% for training and validation sets that ensure a good fit of the classifier model because both sets show an increase of accuracy value for every epoch.

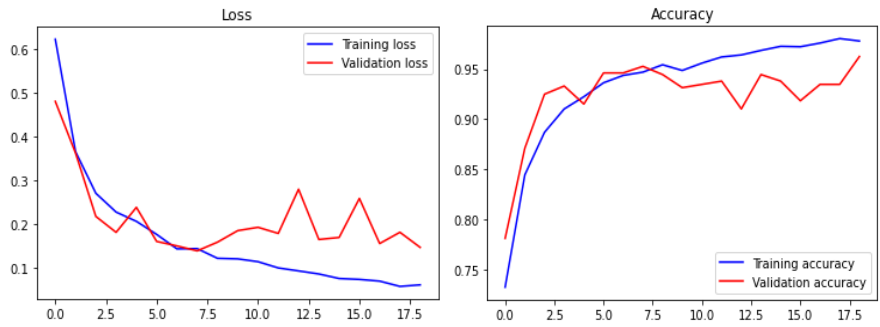

**Fig. 4.** The loss and accuracy of the proposed classification model.

For evaluating the performance of the proposed model, various metrics including precision, recall are obtained at different epochs shown in table 1. The precision and recall of the training set are found 98.16% and 96.39% and for validation set are found 97.55% and 95.92% respectively. Moreover, the F1 Score for training and validation set is calculated 97.85% and 96.15% respectively depending on the following equations.

$$\text{F1Score} = 2\times \frac{precision*Recall}{Pricision+Recall} \times 100\% \tag{1}$$

**Table 1.** Precision and recall of the proposed classifier set at different epochs.

| Epoch | At epoch 1 | At epoch 4 | At epoch 7 | At epoch 10 | At epoch 13 | At epoch 16 | At epoch 19 |
|---|---|---|---|---|---|---|---|
| Training Precision | 0.7859 | 0.9166 | 0.9490 | 0.9524 | 0.9644 | 0.9726 | 0.9816 |
| Validation Precision | 0.8094 | 0.9373 | 0.9522 | 0.9343 | 0.9161 | 0.9195 | 0.9639 |
| Training Recall | 0.6663 | 0.9013 | 0.9412 | 0.9466 | 0.9621 | 0.9718 | 0.9755 |
| Validation Recall | 0.7276 | 0.9266 | 0.9429 | 0.9282 | 0.9086 | 0.9135 | 0.9592 |

Table 2 shows the comparison of the proposed work with some existing research works based on the classification method and performance parameter. From this table, it is shown that the accuracy is quite improved by using the transfer learning based CNN classifier proposed in this work. Therefore, the proposed method is proficient to significantly improve the multi-classification accuracy comparing with other models presented in the table.

Table 2. Comparison of the proposed model with the existing model.

| Reference | Classification Method | Accuracy (%) |
|---|---|---|
| [2] | Modified VGG-19 CNN | 94.58 |
| [4] | CNN | 96.13 |
| [5] | CNN | 94.68 |
| [7] | Deep transfer learning with ResNet-50 | 95.23±0.6 |
| [8] | VGG-16 with LSTM | 84.00 |
| This work | Inception-v3 with customized DNN | **96.25** |

# 4      Conclusion

This paper proposed a method of multi-classification of brain tumor images based on inception-v3 transfer learning with customized DNN in MRI images. The fine tuning transfer learning model is employed after the image augmentation can improve the accuracy of the classifier for precisely classifying the tumor images as glioma, meningioma, and pituitary respectively. The proposed deep neural network based classifier provides satisfactory performance in terms of accuracy of 96.25%. Therefore, the improved classification accuracy has a great impact on the computer aided diagnosis of brain cancer that assists the physicians to make an exact decision for the treatment of the neurological patient.